# *Analytical solution for the stopping power of the Cherenkov radiation in a uniaxial nanowire material*

Tiago A. Morgado, David E. Fernandes, Mário G. Silveirinha[*]

*Department of Electrical Engineering, Instituto de Telecomunicações, University of Coimbra, 3030 Coimbra, Portugal*

*E-mail:* tiago.morgado@co.it.pt, dfernandes@co.it.pt, mario.silveirinha@co.it.pt

## Abstract

We derive closed analytical formulae for the power emitted by moving charged particles in a uniaxial wire medium by means of an eigenfunction expansion. Our analytical expressions demonstrate that in the absence of material dispersion the stopping power of the uniaxial wire medium is proportional to the charges velocity, and that there is no velocity threshold for the Cherenkov emission. It is shown that the eigenfunction expansion formalism can be extended to the case of dispersive lossless media. Furthermore, in presence of material dispersion the optimal charge velocity that maximizes the emitted Cherenkov power may be less than the speed of light in vacuum.

---

[*] To whom correspondence should be addressed: E-mail: mario.silveirinha@co.it.pt

## I.  INTRODUCTION

The Cherenkov effect [1-2] has been a topic of continuous interest and research owing to its many applications [3] particularly in particle detection in high energy physics [4], in the development of light sources [5-7], in spectroscopy of nanostructures [8], amongst others. The Cherenkov effect occurs when a charged particle (e.g. an electron) propagates inside a dielectric medium with a velocity larger than the electromagnetic wave phase velocity $v_{ph} = c/n$ of the medium (*c* is the light speed in vacuum and *n* is the refractive index of the medium). A particle with velocity exceeding such a threshold gives rise to a conical wave front, being the emitted light launched along the forward direction $\theta = \arccos v_{ph}/v$ measured with respect to the particle velocity *v*.

The emission of Cherenkov radiation was experimentally discovered by P. A. Cherenkov in 1934 [1]. Some years later, I. M. Frank and I. E. Tamm formalized a theoretical explanation of Cherenkov's observations [9]. In recent years, somewhat triggered by the interest raised by the theoretical work of Veselago [10], the Cherenkov radiation was also investigated in structured materials (metamaterials) [11-23]. In particular, it was proven [11-12, 14-16] that in media with negative refractive index, the emitted Cherenkov radiation may be directed backward relative to the motion of the particle (reversed Cherenkov effect for which $\theta > 90º$), contrarily to what happens in standard dielectrics ($\theta < 90º$). A similar reversed Cherenkov effect may be also observed in photonic crystals [24].

Interestingly, another anomalous property of Cherenkov radiation made possible by metamaterials is the possibility of having Cherenkov radiation with no threshold for the charged particles velocity [17-18, 24]. This remarkable property may be useful to improve the characteristics of free-electron lasers [17]. Such threshold-free Cherenkov emission occurs, for instance, in nanowire metamaterials formed by periodic arrays of parallel metallic nanorods [18]. Interestingly, a nanowire metamaterial structure enables the generation of nondivergent Cherenkov radiation [19-20], and the enhancement of the amount of emitted radiation [18]. It was numerically demonstrated in [18] that the stopping power (defined as the average energy loss of the particles per unit of path length) of a nanowire metamaterial can be more than two orders of magnitude larger than in natural media.

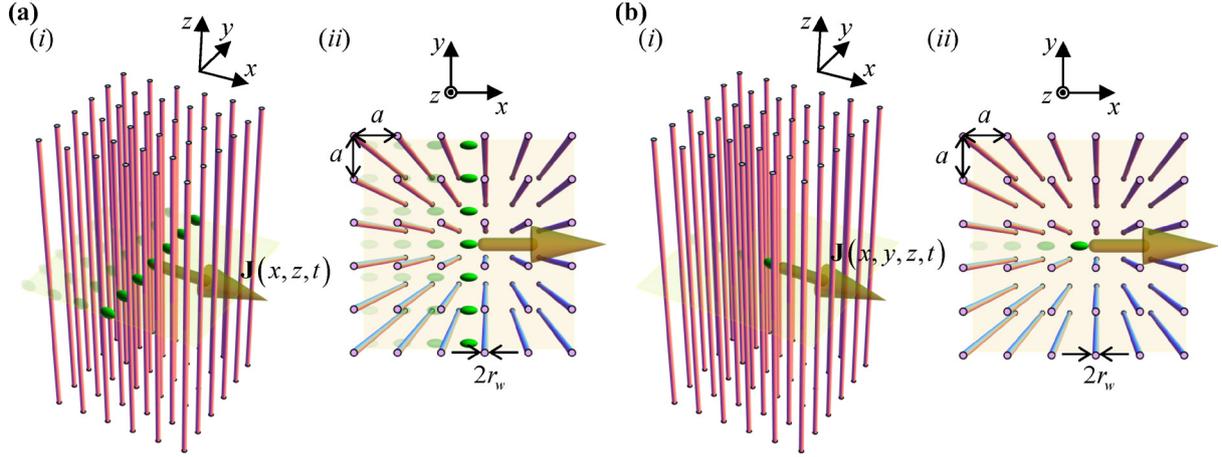

**Fig. 1.** Illustration of the systems under study: charged particles moving along the *x*-direction with a constant velocity *v*, inside an unbounded wire medium formed by metallic wires embedded in vacuum. (a) A linear array of charged particles; (b) a single charged particle. The wires have radius $r_w$ and are arranged in a square lattice with lattice constant *a*.

The objective of this work is to rigorously calculate the power emitted in the form of Cherenkov radiation by charged particles moving inside a nanowire metamaterial. Relying on the quasi-static homogenization framework introduced in [25] and performing a modal expansion, here we derive closed analytical formulae for the stopping power of the nanowire structure. We are interested in two scenarios: (i) a linear array of charged particles [Fig. 1(a)], and (ii) a single charged particle [Fig. 1(b)] moving inside an unbounded wire medium along a direction perpendicular to the nanowires.

This paper is organized as follows. In Sec. II, we characterize the wave dynamics in the uniaxial wire medium based on the quasi-static homogenization approach reported in [25]. Then, in Sec. III we derive closed analytical solutions for the stopping power due to the Cherenkov radiation in the two scenarios illustrated in Fig. 1, assuming that the wires are perfectly electrical conducting (PEC). In Sec. IV, we generalize the theoretical formalism to the case of lossless dispersive media. Finally, in Sec. IV the conclusions are drawn.

Throughout this work we assume that in the case of a time harmonic regime, the time dependence is of the form $e^{-i\omega t}$.

## II. WAVE DYNAMICS IN UNIAXIAL WIRE MEDIA

In this section, we characterize the free oscillation modes of the electromagnetic field in a uniaxial wire medium [Fig. 1] by reducing the problem to the calculation of the spectrum of a Hermitian operator. To this end, we rely on the quasi-static homogenization framework introduced in Ref. [25]. Within this approach, the electromagnetic response of the wire medium is expressed in terms of additional variables with known physical meaning, namely an additional potential $\varphi$ related to the average electric potential drop from a given wire to the boundary of the associated unit cell and an additional current $I$ that represents the electric current flowing along the wire [25]. The wave dynamics in the uniaxial wire medium [Fig. 1] is described by an eight-component state vector $\mathbf{F} = (\mathbf{E}, \mathbf{H}, \varphi, I)^T$, which consists of the macroscopic electric and magnetic fields, the additional potential and the current. Assuming that the wires are PEC, the state vector satisfies a differential system of the form

$$\hat{L} \cdot \mathbf{F} = i\overline{\overline{\mathbf{M}}} \cdot \frac{\partial \mathbf{F}}{\partial t} \qquad (1)$$

where $\hat{L}$ is a first-order linear differential operator (fully independent of the medium response), $\overline{\overline{\mathbf{M}}}$ is an 8×8 material matrix that describes the response of the wire medium and that depends solely on the geometry of the structure and on the electromagnetic properties of the involved materials. The explicit formulas for $\hat{L}$ and $\overline{\overline{\mathbf{M}}}$ can be found in Appendix A.

Let us introduce an inner product $\langle \ | \ \rangle$ such that for two generic eight-component vectors $\mathbf{F}_1$ and $\mathbf{F}_2$ we have

$$\langle \mathbf{F}_2 | \mathbf{F}_1 \rangle = \frac{1}{2V} \int d^3\mathbf{r} \ \mathbf{F}_2^* \cdot \overline{\overline{\mathbf{M}}} \cdot \mathbf{F}_1, \qquad (2)$$

where $V = L_x \times L_y \times L_z$ is the volume of the region of interest. In the end, we take the limit $V \to \infty$. The symbol "*" denotes complex conjugation. Since $\overline{\overline{\mathbf{M}}}$ is positive definite, it is evident that $\langle \mathbf{F} | \mathbf{F} \rangle > 0$ for $\mathbf{F} \neq 0$, and thus $\langle \ | \ \rangle$ really defines an inner product in the space of eight-component vectors. Indeed, $\langle \mathbf{F} | \mathbf{F} \rangle$ has the physical meaning of the stored energy normalized to the volume of the system [26]:

$$\langle \mathbf{F} | \mathbf{F} \rangle = \frac{1}{2V} \int d^3\mathbf{r} \left( \varepsilon_0 \varepsilon_h |\mathbf{E}|^2 + \mu_0 |\mathbf{H}|^2 + \frac{C_w}{A_c} |\varphi|^2 + \frac{L_w}{A_c} |I|^2 \right), \qquad (3)$$

where $\varepsilon_0$ is the electric permittivity of free-space, $\varepsilon_h$ is the relative permittivity of the host medium, $\mu_0$ is the magnetic permeability of free-space, $A_c = a^2$, and $C_w$ and $L_w$ are the effective capacitance and inductance of the wires per unit length of a wire, respectively [25].

Importantly, it may be checked that the operator $\overline{\overline{\mathbf{M}}}^{-1} \cdot \hat{L}$ is a Hermitian operator in the Hilbert space of eight-component vectors that satisfy periodic boundary conditions with an inner product defined as in Eq. (2), i.e. $\langle \mathbf{F}_2 | \overline{\overline{\mathbf{M}}}^{-1} \hat{L} \cdot \mathbf{F}_1 \rangle = \langle \overline{\overline{\mathbf{M}}}^{-1} \hat{L} \cdot \mathbf{F}_2 | \mathbf{F}_1 \rangle$. In particular, it follows that $\overline{\overline{\mathbf{M}}}^{-1} \hat{L}$ has a complete set of eigenfunctions $\mathbf{F}_n = (\mathbf{E}_n, \mathbf{H}_n, \varphi_n, I_n)^T$, such that

$$\overline{\overline{\mathbf{M}}}^{-1} \hat{L} \cdot \mathbf{F}_n = \omega_n \mathbf{F}_n, \qquad (4)$$

where $\omega_n$ are the eigenfrequencies and $n=1,2,\ldots$. The normalization of the modes is chosen so that they are orthonormal, i.e.,

$$\langle \mathbf{F}_m | \mathbf{F}_n \rangle = \frac{1}{2V} \int d^3\mathbf{r} \ \mathbf{F}_m^* \cdot \overline{\overline{\mathbf{M}}} \cdot \mathbf{F}_n = \delta_{m,n}. \qquad (5)$$

Since the set of eigenfunctions $\mathbf{F}_n$ is complete, a generic eight-component vector $\mathbf{F}$ can be expanded as follows

$$\mathbf{F} = \sum_n d_n \mathbf{F}_n, \qquad d_n = \langle \mathbf{F}_n | \mathbf{F} \rangle. \tag{6}$$

The effective medium is invariant to translations, and thus the dependence of the eigenfunctions on the spatial coordinates is of the form $e^{i\mathbf{k}\cdot\mathbf{r}}$. It is evident that for each wave vector $\mathbf{k}$ the eigenvalue problem (4) reduces to a standard 8×8 matrix eigensystem. Hence, there are eight different eigenwaves (eigenfunctions). Consistent with [27], it is found that the eigenfunctions split into the following classes: transverse electric (TE) waves, transverse magnetic (TM) waves, and transverse electromagnetic (TEM) waves. For a fixed $\mathbf{k}$ the eigenvalue problem (4) has exactly two solutions of each type. Because the system is invariant under a composition of a parity transformation ($\mathcal{P}$) and the complex conjugation operation ($\mathcal{K}$), the two eigenmodes of the same type differ by the $\mathcal{KP}$ symmetry transformation. The eigenfrequency $\omega_n$ of a given mode is transformed as $\omega_n \to -\omega_n$ under the $\mathcal{KP}$ operation. In addition, for a fixed $\mathbf{k}$ the eigenvalue problem (4) also supports two longitudinal static (LS) (electrostatic and magnetostatic) modes associated with $\omega_n = 0$.

### III.  STOPPING POWER OF UNIAXIAL WIRE MEDIA

Next, the theoretical formalism of Sec. II is used to obtain the power emitted due to the Cherenkov radiation by charged particles moving inside a uniaxial wire medium. In presence of an external source the electrodynamics of the problem is described by the system of equations:

$$\hat{L}\mathbf{F} = i\overline{\overline{\mathbf{M}}} \cdot \frac{\partial \mathbf{F}}{\partial t} + i\mathbf{J}_{ext}, \tag{7}$$

where $\mathbf{J}_{ext}$ is an eight-component vector given by $\mathbf{J}_{ext} = (\mathbf{j}_{ext}, \mathbf{0}, 0, 0)^T$, where $\mathbf{j}_{ext}$ is the electric current density and $\mathbf{0}$ is the zero vector. In the frequency domain Eq. (7) becomes

$$\hat{L}\mathbf{F}_\omega = \omega \overline{\overline{\mathbf{M}}} \cdot \mathbf{F}_\omega + i\mathbf{J}_{ext,\omega},\tag{8}$$

where $\mathbf{F}_\omega$ and $\mathbf{J}_{ext,\omega}$ are Fourier transforms in time. Expanding $\mathbf{F}_\omega$ as in Eq. (6) and taking into account that $\hat{L}\mathbf{F}_\omega = \sum_n d_n \omega_n \overline{\overline{\mathbf{M}}} \cdot \mathbf{F}_n$ it is found that (8) is equivalent to:

$$\sum_n d_n (\omega_n - \omega) \mathbf{F}_n = \overline{\overline{\mathbf{M}}}^{-1} \cdot i\mathbf{J}_{ext,\omega}.\tag{9}$$

Thus, because of the orthogonality conditions (5) the coefficients $d_n$ must satisfy:

$$d_n = \frac{1}{\omega_n - \omega} \left\langle \mathbf{F}_n \middle| \overline{\overline{\mathbf{M}}}^{-1} \cdot i\mathbf{J}_{ext,\omega} \right\rangle.\tag{10}$$

Thus, we finally conclude that:

$$\mathbf{F}_\omega(\mathbf{r}) = \sum_n \mathbf{F}_n(\mathbf{r}) \frac{1}{\omega_n - \omega} \left\langle \mathbf{F}_n \middle| \overline{\overline{\mathbf{M}}}^{-1} \cdot i\mathbf{J}_{ext,\omega} \right\rangle.\tag{11}$$

The stopping power is given by $P_0 / v$, where $P_0$ is the total instantaneous power extracted from charged particles moving at a constant velocity [2]. Specifically, $P_0 = -\int d^3\mathbf{r}\, \mathbf{E} \cdot \mathbf{j}_{ext}$, being $\mathbf{E}$ the total electric field that acts on the charged particles. It should be noted that as the charged particles are not accelerated, the self-field does not contribute to the stopping power [2]. Within the eight-component vector notation, $P_0$ can be expressed as

$$P_0 = -\int d^3\mathbf{r}\, \mathbf{F}(\mathbf{r},t) \cdot \mathbf{J}_{ext}(\mathbf{r},t).\tag{12}$$

In what follows, we obtain analytical expressions for the stopping power in the two scenarios illustrated in Fig. 1.

### A. Array of charges moving inside the wire medium

Here, we consider the situation wherein a linear array of charged particles moves inside an unbounded wire medium with a constant velocity *v* and along a direction perpendicular to the wires [Fig. 1(a)]. Supposing that the motion is confined to the $z = z_0$ plane, the current density of the moving charges may be written as $\mathbf{j}_{ext} = -e n_y v \delta(z-z_0) \delta(x-vt) \hat{\mathbf{x}}$, where $n_y$ is the number of charges per unit of length along the y-direction and $-e$ is the electron charge. In this case, $\mathbf{J}_{ext,\omega} = -e n_y \delta(z-z_0) e^{i\frac{\omega}{v}x} \hat{\mathbf{u}}_s$, where $\hat{\mathbf{u}}_s = (\hat{\mathbf{x}}, \mathbf{0}, 0, 0)^T$ should be understood as a eight-component unit vector. Then, it is straightforward to check that Eq. (11) reduces to:

$$\mathbf{F}_\omega(\mathbf{r}) = -i e n_y \sum_n \mathbf{F}_n(\mathbf{r}) \frac{1}{\omega_n - \omega} \frac{1}{2V} \int dx' dy' \, \mathbf{F}_n^*(x',y',z_0) \cdot \hat{\mathbf{u}}_s e^{i\frac{\omega}{v}x'}. \tag{13}$$

Next, we write the eigenmodes $\mathbf{F}_n$ in the form $\mathbf{F}_n = \mathbf{F}_{m\mathbf{k}} = \mathbf{F}_{m\mathbf{k},0} e^{i\mathbf{k}\cdot\mathbf{r}}$, where $\mathbf{F}_{m\mathbf{k},0}$ is a constant vector independent of $\mathbf{r}$, $\mathbf{k}$ is the wave vector associated with the eigenmode, and the index $m$= TE, TM, TEM or LS, determines the electromagnetic mode type. In the continuous limit ($V \to \infty$), the discrete summation in Eq. (13) must be replaced by an integration over $\mathbf{k}$ such that $\frac{1}{V}\sum_n \to \frac{1}{(2\pi)^3}\sum_m \int d^3\mathbf{k}$. Hence, it follows that:

$$\mathbf{F}_\omega(\mathbf{r}) = -i e n_y \frac{1}{(2\pi)^3} \int d^3\mathbf{k} \sum_m \mathbf{F}_{m\mathbf{k}}(\mathbf{r}) \frac{\mathbf{F}_{m\mathbf{k},0}^* \cdot \hat{\mathbf{u}}_s}{\omega_{m\mathbf{k}} - \omega} \frac{1}{2} \int dx' dy' \, e^{-i(k_x x' + k_y y' + k_z z_0)} e^{i\frac{\omega}{v}x'}. \tag{14}$$

where $\omega_{m\mathbf{k}}$ are the resonant frequencies associated with the Floquet eigenmodes with wave vector $\mathbf{k}$. Straightforward simplifications give the final result for $\mathbf{F}_\omega$:

$$\begin{aligned}\mathbf{F}_\omega(\mathbf{r}) &= -i e n_y \frac{1}{4\pi} \sum_m \int dk_z dk_x \left( \frac{1}{\omega_{m\mathbf{k}} - \omega} \mathbf{F}_{m\mathbf{k}}(\mathbf{r}) \mathbf{F}_{m\mathbf{k},0}^* \cdot \hat{\mathbf{u}}_s e^{-ik_z z_0} \delta\left(k_x - \frac{\omega}{v}\right)\right)\bigg|_{k_y=0} \\ &= -i e n_y \frac{1}{4\pi} \sum_m \int dk_z \left( \frac{1}{\omega_{m\mathbf{k}} - \omega} \mathbf{F}_{m\mathbf{k}}(\mathbf{r}) \mathbf{F}_{m\mathbf{k},0}^* \cdot \hat{\mathbf{u}}_s e^{-ik_z z_0} \right)\bigg|_{\substack{k_x=\omega/v \\ k_y=0}}\end{aligned}. \tag{15}$$

Next, we calculate the inverse Fourier transform to obtain $\mathbf{F}$ in the time domain. As usual, the singularity of the integrand is avoided by replacing $\omega \to \omega + 0i$ (integration path is in the upper half plane, consistent with the causality of the system response). Using again $\mathbf{F}_{m\mathbf{k}} = \mathbf{F}_{m\mathbf{k},0} e^{i\mathbf{k}\cdot\mathbf{r}}$, it is found that:

$$\mathbf{F}(\mathbf{r},t) = -ien_y \frac{|v|}{8\pi^2} \sum_m \int dk_z dk_x \left( \frac{1}{\omega_{m\mathbf{k}} - (k_x v + 0^+ i)} e^{ik_x(x-vt)} e^{ik_z(z-z_0)} \mathbf{F}_{m\mathbf{k},0} \mathbf{F}^*_{m\mathbf{k},0} \cdot \hat{\mathbf{u}}_s \right)\bigg|_{k_y=0}. \quad (16)$$

We are now ready to determine the stopping power of the nanowire metamaterial. Substituting the above formula into Eq. (12) and using $\mathbf{J}_{\text{ext}}(x,z,t) = -en_y \delta(z-z_0)\delta\left(\frac{x}{v}-t\right)\hat{\mathbf{u}}_s$ one obtains:

$$\frac{P_0}{L_y} = \frac{(en_y v)^2}{8\pi^2} \sum_m \int dk_z dk_x \left( \frac{-i}{\omega_{m\mathbf{k}} - (k_x v + 0^+ i)} |\mathbf{F}_{m\mathbf{k},0} \cdot \hat{\mathbf{u}}_s|^2 \right)\bigg|_{k_y=0}, \quad (17)$$

where $L_y$ represents the width of the array of charged particles along the $y$-direction, so that $N_y = n_y L_y$ is the total number of moving charges. Using the identity [28]:

$$\frac{1}{x - 0^+ i} = \text{P.V.} \frac{1}{x} + i\pi\delta(x), \quad (18)$$

where P.V. stands for the Cauchy principal value, we may rewrite Eq. (17) as follows:

$$\frac{P_0}{L_y} = \frac{(en_y v)^2}{8\pi^2} \sum_m \int dk_z dk_x \, \text{P.V.} \left( \frac{-i}{\omega_{m\mathbf{k}} - k_x v} |\mathbf{F}_{m\mathbf{k},0} \cdot \hat{\mathbf{u}}_s|^2 \right)\bigg|_{k_y=0} + \\ + \frac{(en_y v)^2}{8\pi} \sum_m \int dk_z dk_x \left( \delta(\omega_{m\mathbf{k}} - k_x v) |\mathbf{F}_{m\mathbf{k},0} \cdot \hat{\mathbf{u}}_s|^2 \right)\bigg|_{k_y=0}. \quad (19)$$

The first term is pure imaginary and hence the corresponding integral must vanish. Therefore, Eq. (19) becomes simply

$$\frac{P_0}{L_y} = \frac{(en_y v)^2}{8\pi} \sum_m \int dk_z dk_x \left( \delta(\omega_{m\mathbf{k}} - k_x v) |\mathbf{F}_{m\mathbf{k},0} \cdot \hat{\mathbf{u}}_s|^2 \right)\bigg|_{k_y=0}. \quad (20)$$

This formula shows that the natural modes that contribute to the Cherenkov radiation satisfy the selection rules $k_x = \frac{\omega_{m\mathbf{k}}}{v}$ and $k_y = 0$. This is consistent with the fact that in the frequency domain the excitation varies with $x$ and $y$ as $\mathbf{J}_{\text{ext},\omega} \sim e^{i\frac{\omega}{v}x}$. It is important to emphasize that there are natural modes with $\omega_{m\mathbf{k}}$ positive, and modes with $\omega_{m\mathbf{k}}$ negative, and that the two set of modes are related by the $\mathcal{KP}$ transformation. As previously mentioned, the index $m$ identifies the electromagnetic mode type (TE, TM, TEM or LS). It is easy to check that the TE and LS modes are not excited in the scenario of Fig. 1(a) because $\mathbf{F}_{m\mathbf{k},0} \cdot \hat{\mathbf{u}}_s = 0$. Thus, the sum over $m$ may be restricted to TEM and TM modes.

The main radiative channel in the uniaxial wire medium is associated with the TEM eigenmode [18]. This mode has no cutoff frequency and is the only one that propagates below the plasma frequency of the effective medium ($\omega_p = \beta_p c$, with $\beta_p a = \sqrt{\mu_0/L_w} = \sqrt{2\pi/(\ln(a/(2\pi r_w))+0.5275)/a}$) [25, 27, 29]. The dispersion characteristic of the (positive and negative frequency) TEM eigenmodes is given by $\omega = \pm|k_z|c_h$, where $c_h = c/\sqrt{\varepsilon_h}$. Then, from Eq. (20) it is possible to write:

$$\frac{P_0}{L_y} = \frac{(en_y v)^2}{8\pi} \int dk_z dk_x \left[ \delta(|k_z|c_h - k_x v) + \delta(-|k_z|c_h - k_x v) \right] \left|\mathbf{E}_{\text{TEM},\mathbf{k},0} \cdot \hat{\mathbf{x}}\right|^2 \bigg|_{k_y=0}$$
$$= \frac{2(en_y v)^2}{8\pi c_h} \int_{-\infty}^{+\infty} dk_x \left|\mathbf{E}_{\text{TEM},\mathbf{k},0} \cdot \hat{\mathbf{x}}\right|^2 \bigg|_{k_y=0} \qquad (21)$$

The electric field associated with the TEM modes is of the form $\mathbf{E}_{\text{TEM},\mathbf{k},0} \sim A\mathbf{k}_\parallel$, where $A$ is a normalization constant such that $\langle \mathbf{F}|\mathbf{F}\rangle = 1$ and $\mathbf{k}_\parallel = \mathbf{k} - \mathbf{k}\cdot\hat{\mathbf{z}}\hat{\mathbf{z}}$ is the transverse component of the wave vector. Using Eqs. (A1)-(A4) together with Eq. (3) one can prove that

$$|A|^2 = \frac{\beta_p^2}{(k_x^2+k_y^2)(k_x^2+k_y^2+\beta_p^2)\varepsilon_0\varepsilon_h} \text{ so that}$$

$$\left|\mathbf{E}_{\text{TEM},\mathbf{k},0} \cdot \hat{\mathbf{x}}\right|^2 = \frac{k_x^2 \beta_p^2}{(k_x^2+k_y^2)(k_x^2+k_y^2+\beta_p^2)\varepsilon_0\varepsilon_h}. \qquad (22)$$

Substituting this formula into Eq. (21) with $k_y = 0$, and calculating the integral in $k_x$ over the first Brillouin zone ($[-\pi/a, \pi/a]$) [31], it is finally found that:

$$\frac{P_0}{v} = \frac{1}{2\pi} \frac{v}{c} \frac{e^2 n_y N_y}{\varepsilon_0} \frac{\beta_p}{\sqrt{\varepsilon_h}} \arctan\left(\frac{\pi}{a\beta_p}\right), \qquad (23)$$

where $N_y = n_y L_y$ is the total number of moving charges. Thus, the contribution of the TEM mode to the stopping power is always nonzero and increases with the velocity $v$. The integration was restricted to the Brillouin zone because the effective medium theory breaks down when $|k_x| > \pi/a$ [30-31]. Indeed, in the framework of a microscopic theory the TEM waves have always a transverse wave vector confined to the first Brillouin zone. Our theory is expected to capture accurately the physics of the Cherenkov effect when the moving charges do not excite spatial harmonics with $|k_x| > \pi/a$. This requires that the electron beam longitudinal width (along the direction of motion) is larger than the period $a$ of the nanowire material. Hence, the $\delta$-function in the definition of $\mathbf{j}_\text{ext}$ should be understood as a function peaked at the origin with a width of the order of the lattice constant of the wire medium.

In Fig. 2, we compare the analytical formula (23) (solid lines) with the numerical results calculated with the theory of Ref. [18] (discrete symbols). An excellent agreement between the two formalisms is revealed, supporting in this manner the validity of Eq. (23).

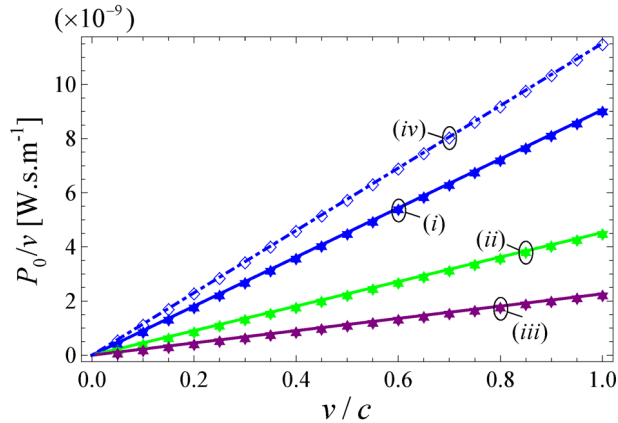

**Fig. 2.** Stopping power ($P_0/v$) as a function of the normalized velocity ($v/c$) for a linear array of charged particles moving inside the uniaxial wire medium [Fig. 1(a)] formed by PEC wires, $\varepsilon_h = 1$, $n_y = 100\ \mu\text{m}^{-1}$, $N_y = 20 \times 10^3$, and different values of the lattice spacing $a$ and radius of the wires $r_w$. (i) $a = 200$ nm and $r_w = 0.05a$; (ii) $a = 400$ nm and $r_w = 0.05a$; (iii) $a = 800$ nm and $r_w = 0.05a$; (iv) $a = 200$ nm and $r_w = 0.15a$. Solid lines: results obtained from the analytical formula (23); Discrete symbols: numerical results based on the theory of [18].

The property that stands out from Fig. 2 is the fact that there is no threshold for the velocity of the moving charged particles. Therefore, unlike usual dielectric materials, the uniaxial wire medium allows extracting power from the charges even when they are moving at relatively low velocities.

It can be also seen from Fig. 2 that, as the separation between the wires $a$ is reduced, the magnitude of the stopping power increases. In fact, this could be expected from the analytical formula (23) because the plasma frequency is inversely proportional to the lattice spacing, $\beta_p \sim 1/a$. Such an enhancement of the stopping power occurs because the photonic states density increases with the density of wires [18], leading to a boost of the number of available radiative channels. On the other hand, Fig. 2 shows that the value of the stopping power also increases with the radius of the wires. This happens because the coupling between the charges and the wires becomes stronger for larger radii.

Even though the main radiative channel in the uniaxial wire medium is related to the TEM mode, it is known [18] that the TM mode can provide a secondary radiative channel and thereby also contribute to the Cherenkov emission. However, the TM radiative channel only becomes available for velocities greater than the threshold $v > c/\sqrt{\varepsilon_h}$ [18]. In particular, when the host medium is a vacuum ($\varepsilon_h = 1$), as considered in Fig. 2, the TM mode does not contribute to the stopping power, and Eq. (23) is exact. It may be checked that when $\varepsilon_h > 1$ and the host material dispersion is ignored the contribution of the TM mode to the stopping power is infinitely large. Thus, if the host is not a vacuum it is essential to include the effects of material dispersion in the calculation of the stopping power. We discuss how the material dispersion can be taken into account in Sec. IV.

### B. Single charge moving inside the wire medium

In what follows, we extend the study of Sec. III-A to the case wherein a single charged particle moves inside the nanowire structure along a straight line with constant $y$ and $z$ (namely, $y = y_0$ and $z = z_0$) [Fig. 1(b)]. In this scenario, the current density may be written as $\mathbf{j}_{ext} = -ev\delta(z-z_0)\delta(y-y_0)\delta(x-vt)\hat{\mathbf{x}}$, and hence in the frequency domain $\mathbf{J}_{ext,\omega} = -e\delta(z-z_0)\delta(y-y_0)e^{i\frac{\omega}{v}x}\hat{\mathbf{u}}_s$. Using again Eq. (11) it is found after some simplifications that:

$$\mathbf{F}_\omega(\mathbf{r}) = -ie\frac{1}{8\pi^2}\sum_m \int d^3k \frac{1}{\omega_{m\mathbf{k}}-(\omega+0^+i)}\delta\left(k_x - \frac{\omega}{v}\right)e^{-ik_y y_0}e^{-ik_z z_0}\mathbf{F}_{m\mathbf{k}}(\mathbf{r})\mathbf{F}^*_{m\mathbf{k},0}\cdot\hat{\mathbf{u}}_s. \quad (24)$$

Calculating the inverse Fourier transform in time, we obtain:

$$\mathbf{F}(\mathbf{r},t) = -\frac{ie|v|}{16\pi^3}\sum_m \int d^3k \frac{1}{\omega_{m\mathbf{k}}-(k_x v+0^+i)}e^{ik_x(x-vt)}e^{ik_y(y-y_0)}e^{ik_z(z-z_0)}\mathbf{F}_{m\mathbf{k},0}\mathbf{F}^*_{m\mathbf{k},0}\cdot\hat{\mathbf{u}}_s. \quad (25)$$

Substituting this result into Eq. (12) and using the relation (18), we obtain the total instantaneous power extracted from the moving charge:

$$P_0 = \frac{(ev)^2}{16\pi^2}\sum_m \int d^3k\, \delta(\omega_{m\mathbf{k}}-k_x v)\left|\mathbf{F}_{m\mathbf{k},0}\cdot\hat{\mathbf{u}}_s\right|^2. \quad (26)$$

Hence, the natural modes that contribute to the Cherenkov radiation by a single moving charge satisfy a single selection rule $k_x = \frac{\omega_{m\mathbf{k}}}{v}$. When the nanowires stand in a vacuum ($\varepsilon_h = 1$) only the TEM modes can satisfy this selection rule, because for TE and TM waves $|k_x| \leq \frac{\omega}{c} < \frac{\omega}{v}$ while for LS waves $\mathbf{F}_{m\mathbf{k},0}\cdot\hat{\mathbf{u}}_s = 0$ when $k_x = 0$. Hence, in what follows we restrict our attention to the TEM waves contribution to Cherenkov emission. Proceeding as in the previous subsection, and using Eq. (22) it is found that:

$$P_0 = \frac{(ev)^2}{8\pi^2 c_h} \iint dk_x dk_y \left| \mathbf{E}_{\text{TEM},\mathbf{k},0} \cdot \hat{\mathbf{x}} \right|^2 \bigg|_{k_z = \pm k_x \frac{v}{c_h}}$$
$$= \frac{(ev)^2}{8\pi^2 c_h} \iint dk_x dk_y \frac{k_x^2 \beta_p^2}{(k_x^2 + k_y^2)(k_x^2 + k_y^2 + \beta_p^2)\varepsilon_0 \varepsilon_h}. \quad (27)$$

The double integral is divergent if the integration range is taken to be all space. However, as previously discussed, it is known that the values of $(k_x, k_y)$ for a TEM wave are required to lie within the first Brillouin zone ($[-\pi/a, \pi/a] \times [-\pi/a, \pi/a]$) [31]. For convenience, we approximate the square shaped Brillouin zone by a circular region ($k < k_{\max}$) with the same area ($k_{\max} = 2\sqrt{\pi}/a$). Then, the integral (27) may be written in cylindrical coordinates as follows:

$$P_0 = \frac{(ev)^2}{8\pi^2 c_h} \int_0^{k_{\max}} \int_0^{2\pi} dk d\varphi \, k \frac{\cos^2 \varphi \, \beta_p^2}{(k^2 + \beta_p^2)\varepsilon_0 \varepsilon_h}. \quad (28)$$

Thus, the stopping power for a single moving charge is:

$$\frac{P_0}{v} = \frac{v}{16\pi c} \frac{e^2}{\varepsilon_0} \frac{\beta_p^2}{\sqrt{\varepsilon_h}} \ln\left(\frac{4\pi}{\beta_p^2 a^2} + 1\right). \quad (29)$$

Therefore, analogous to what happens in the scenario wherein an array of charges moves inside the nanowire material [see Eq. (23)], the stopping power is also here an increasing linear function of the velocity $v$ of the charged particles.

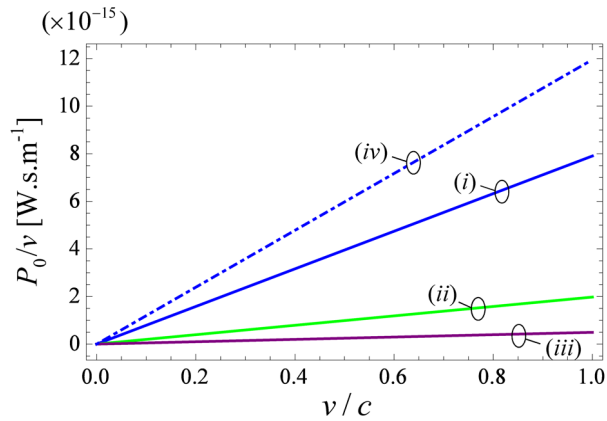

**Fig. 3.** Stopping power ($P_0/v$) as a function of the normalized velocity ($v/c$) for a single charged particle moving inside the uniaxial wire medium [Fig. 1(b)] formed by PEC wires, $\varepsilon_h = 1$, and different values of the lattice spacing $a$ and radius of the wires $r_w$, calculated using the closed-analytical formula (29). (*i*) $a = 200$ nm and $r_w = 0.05a$; (*ii*) $a = 400$ nm and $r_w = 0.05a$; (*iii*) $a = 800$ nm and $r_w = 0.05a$; (*iv*) $a = 200$ nm and $r_w = 0.15a$.

In Fig. 3 we depict the dependence of the stopping power on the velocity $v$ of the charged particle for different structural parameters, calculated using Eq. (29). Similar to the case of a linear array of moving charges, it is seen that the stopping power increases as the separation between the wires $a$ is reduced or the radius of the wires is enlarged.

It is interesting to note that the power extracted per charge in the case of the linear array is given by $\dfrac{P_0}{N_y} = \dfrac{1}{2\pi} \dfrac{v^2}{c} \dfrac{e^2 n_y}{\varepsilon_0} \dfrac{\beta_p}{\sqrt{\varepsilon_h}} \arctan\left(\dfrac{\pi}{a\beta_p}\right)$ which differs from the power extracted by a single moving charge. Indeed, we have that:

$$\frac{P_0|_{\text{single charge}}}{\dfrac{P_0}{N_y}\bigg|_{\text{linear array}}} = \frac{1}{8} \frac{\beta_p}{n_y} \frac{\ln\left(\dfrac{4\pi}{\beta_p^2 a^2} + 1\right)}{\arctan\left(\dfrac{\pi}{a\beta_p}\right)}. \tag{30}$$

In particular, if the number of charges per cell is large one has $\beta_p/n_y \ll 1$ and consequently $P_0|_{\text{single charge}} \ll \dfrac{P_0}{N_y}\bigg|_{\text{linear array}}$. Therefore, one sees that the interference between the fields emitted by a linear array of moving charges contributes to enhance the stopping power. The physical justification is that the interference of emitted fields suppresses radiation channels with $k_y \neq 0$, and promotes the emission into the *xoz* plane which is a more efficient process ($|\mathbf{E}_{\text{TEM},\mathbf{k},0} \cdot \hat{\mathbf{x}}|^2$ is maximum in the *xoz* plane).

## IV.    GENERALIZATION TO DISPERSIVE MEDIA

The analysis of Sec. III assumes that both the host medium and the nanowires are dispersionless. However, in practice the permittivity of realistic materials depends on frequency. In general, the effects of material dispersion are essential to obtain a finite emitted power. Next, we explain how the theory can be generalized in a straightforward manner to the case of lossless dispersive media.

It is well known that one can model the electromagnetic response of lossless dispersive dielectrics and metals using a Hermitian formulation (see [32]). Using such an approach, it is possible to describe the wave dynamics as in Eq. (7), but with different operators $\hat{L}$ and $\overline{\overline{\mathbf{M}}}$ and for an extended state vector $\mathbf{F} = (\mathbf{E}, \mathbf{H}, \varphi, I, ...)^T$ with $8+M$ components, where the additional $M$ variables describe the internal degrees of freedom of the pertinent dielectrics and metals. Based on such a formulation, it is possible to repeat all the steps of Sec. III and prove that the stopping power for the configuration of Fig. 1(a) is still given by Eq. (20), which is equivalent to:

$$\frac{P_0}{L_y} = \frac{(en_y v)^2}{8\pi} \sum_m \int dk_z dk_x \left( \delta(\omega_{m\mathbf{k}} - k_x v) |\mathbf{E}_{m\mathbf{k},0} \cdot \hat{\mathbf{x}}|^2 \right)\bigg|_{k_y=0}. \tag{31}$$

Performing the integration in $k_z$ it is found that:

$$\frac{P_0}{L_y} = 2 \times \frac{(en_y v)^2}{8\pi} \sum_{\substack{\text{Im}\{k_z^{(m)}\}=0,\\ \omega_{m\mathbf{k}}>0}} \int dk_x \left( \frac{1}{|\partial_{k_z}\omega_{m\mathbf{k}}|} |\mathbf{E}_{m\mathbf{k},0} \cdot \hat{\mathbf{x}}|^2 \right)\bigg|_{k_y=0, k_z=k_z^{(m)}(k_x)>0}. \tag{32}$$

where $k_z^{(m)}(k_x)$ is obtained by solving $\omega_{m\mathbf{k}} - k_x v = 0$ with respect to $k_z$. Only modes that satisfy $\text{Im}\{k_z^{(m)}\} = 0$ contribute to the emitted power. The leading factor of 2 is due to the fact that we only consider branches with $k_z^{(m)} > 0$. The sum is restricted to modes with positive frequencies because $\omega_{m\mathbf{k}} - k_x v = 0$ cannot be simultaneously satisfied by two modes that differ by the $\mathcal{KP}$ transformation. As in previous examples, the integration in $k_x$ is restricted to the first Brillouin zone.

Of course, in presence of dispersive materials the dispersion of the eigenwaves $\omega_{m\mathbf{k}}$ is different from what was considered in Sec. III. The simplest way to determine $\omega_{m\mathbf{k}}$ is using the effective medium dielectric function $\overline{\overline{\varepsilon}}(\omega,\mathbf{k})$ of the nanowire material, which can be calculated as detailed in [33-34]. Similarly, $\mathbf{E}_{m\mathbf{k},0}$ can also be determined using the effective medium dielectric function $\overline{\overline{\varepsilon}}(\omega,\mathbf{k})$, apart from a multiplicative constant $A$. Now the challenge is to determine the value of the multiplicative constant $A$. Indeed, $\mathbf{E}_{m\mathbf{k},0}$ is formed by a subset of elements of the extended eigenvector $\mathbf{F}_{m\mathbf{k}}$ which should be normalized such that $\langle \mathbf{F}_{m\mathbf{k}} | \mathbf{F}_{m\mathbf{k}} \rangle = 1$. Thus, it may seem that one needs to know the explicit expressions of operators $\hat{L}$ and $\overline{\overline{\mathbf{M}}}$ to determine the unknown multiplicative constant. Fortunately, this is not so, and it is possible to find $\mathbf{E}_{m\mathbf{k},0}$ based uniquely on the dielectric function $\overline{\overline{\varepsilon}}(\omega,\mathbf{k})$, without knowing the explicit forms of $\hat{L}$ and $\overline{\overline{\mathbf{M}}}$.

The key observation is that $\langle \mathbf{F}_{m\mathbf{k}} | \mathbf{F}_{m\mathbf{k}} \rangle$ is precisely the stored electromagnetic energy of the system (per unit of volume), independent of the number of internal degrees of freedom of the pertinent dielectrics and metals. To prove this, we note that Eq. (7) implies that:

$$-i \left\langle \mathbf{F} | \overline{\overline{\mathbf{M}}}^{-1} \hat{L} | \mathbf{F} \right\rangle = \left\langle \mathbf{F} | \frac{\partial \mathbf{F}}{\partial t} \right\rangle + \left\langle \mathbf{F} | \overline{\overline{\mathbf{M}}}^{-1} \cdot \mathbf{J}_{\text{ext}} \right\rangle, \quad (33)$$

where the inner product is defined as in Eq. (2). Because the operator $\overline{\overline{\mathbf{M}}}^{-1} \hat{L}$ is Hermitian with respect to the considered inner product, it follows that:

$$\operatorname{Re}\left\{ \left\langle \mathbf{F} | \frac{\partial \mathbf{F}}{\partial t} \right\rangle \right\} = \operatorname{Re}\left\{ -\left\langle \mathbf{F} | \overline{\overline{\mathbf{M}}}^{-1} \cdot \mathbf{J}_{\text{ext}} \right\rangle \right\}. \quad (34)$$

Therefore, it is possible to write:

$$\frac{d}{dt}\langle \mathbf{F} | \mathbf{F} \rangle = -\frac{1}{V}\int d^3\mathbf{r}\, \text{Re}\{\mathbf{F}^* \cdot \mathbf{J}_{ext}\}. \tag{35}$$

Noting that the extended excitation vector with $8+M$ components must be of the form $\mathbf{J}_{ext} = (\mathbf{j}_{ext}, \mathbf{0}, 0, 0, ....)^T$ we finally conclude that:

$$\frac{d}{dt}\left[\langle \mathbf{F} | \mathbf{F} \rangle V\right] = -\int d^3\mathbf{r}\, \text{Re}\{\mathbf{E}^* \cdot \mathbf{j}_{ext}\}. \tag{36}$$

The right-hand side of the above equation is precisely the power pumped into the system by the external electric current, and hence for a lossless system the left-hand side must be the time rate of the stored electromagnetic energy. Therefore, the stored electromagnetic energy is exactly $\langle \mathbf{F} | \mathbf{F} \rangle V$.

It is well known that the stored energy density in a medium described by the dielectric function $\bar{\bar{\varepsilon}}(\omega, \mathbf{k})$ for a complex natural mode with a space-time variation $e^{-i\omega t}e^{i\mathbf{k}\cdot\mathbf{r}}$ is [2, 35-36]

$$W_{em} = \frac{1}{2}\mathbf{E}^* \cdot \frac{\partial}{\partial \omega}\left[\omega\bar{\bar{\varepsilon}}(\omega, \mathbf{k})\right] \cdot \mathbf{E} + \frac{1}{2}\mu_0 \mathbf{H}^* \cdot \mathbf{H}. \tag{37}$$

This demonstrates that the normalization condition $\langle \mathbf{F}_{m\mathbf{k}} | \mathbf{F}_{m\mathbf{k}} \rangle = 1$ is equivalent to:

$$1 = \frac{1}{2}\mathbf{E}^*_{m\mathbf{k}} \cdot \frac{\partial}{\partial \omega}\left[\omega\bar{\bar{\varepsilon}}(\omega, \mathbf{k})\right] \cdot \mathbf{E}_{m\mathbf{k}} + \frac{1}{2}\mu_0 \mathbf{H}^*_{m\mathbf{k}} \cdot \mathbf{H}_{m\mathbf{k}}. \tag{38}$$

In order to validate this generalized formulation, we consider now a wire medium formed by silver (Ag) nanowires. It is assumed that Ag follows the lossless Drude dispersion model $\varepsilon_m = 1 - \omega_m^2/\omega^2$, where $\omega_m$ is the plasma frequency of the material ($\omega_m/(2\pi) = 2175$ THz) [37]. The wire medium is characterized using the effective medium model described in Refs. [18, 33-34]. Similar to Sec. III-A, in this case the Cherenkov emission is only determined by the quasi-TEM mode.

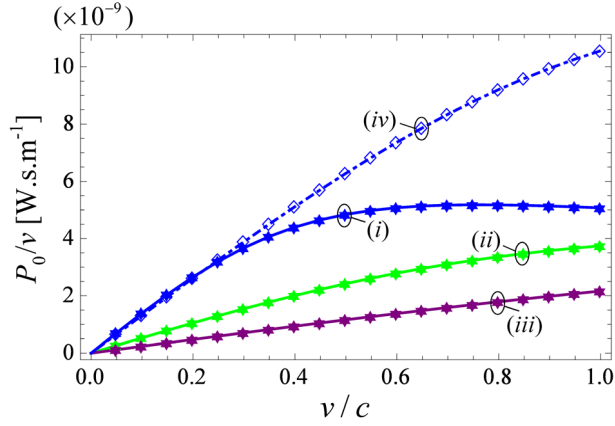

**Fig. 4.** Stopping power ($P_0/v$) as a function of the normalized velocity ($v/c$) for a linear array of charged particles moving inside the uniaxial wire medium [Fig. 1(a)] formed by Ag wires, $\varepsilon_h = 1$, $n_y = 100\ \mu m^{-1}$, $N_y = 20 \times 10^3$, and different values of the lattice spacing $a$ and radius of the wires $r_w$. (*i*) $a = 200$ nm and $r_w = 0.05a$; (*ii*) $a = 400$ nm and $r_w = 0.05a$; (*iii*) $a = 800$ nm and $r_w = 0.05a$; (*iv*) $a = 200$ nm and $r_w = 0.15a$. Solid lines: results obtained from the analytical formula (23); Discrete symbols: numerical results based on the theory of [18].

In Fig. 4, we show that the results obtained with our generalized theory and the numerical results of Ref. [18] (discrete symbols) agree perfectly. It is important to highlight that in presence of material dispersion the stopping power may not vary monotonically with the velocity (curve (*i*)), and that the optimal velocity value may be less than *c* (for curve (*i*) the optimal velocity is $v = 0.75c$).

In summary, we have demonstrated that the stopping power of a nanowire material formed by arbitrary dispersive lossless metals and dielectrics can be computed using Eq. (32) where $\omega_{m\mathbf{k}}$ and $\mathbf{E}_{m\mathbf{k},0}$ are completely characterized by the effective medium dielectric function $\bar{\bar{\varepsilon}}(\omega, \mathbf{k})$, independent of the specific dispersive models that characterize the dielectrics and the metal. The electric field $\mathbf{E}_{m\mathbf{k},0}$ is normalized so that Eq. (38) is satisfied.

## V. CONCLUSION

In this work, we derived closed analytical expressions for the stopping power associated with the Cherenkov emission of charged particles moving inside a uniaxial wire medium formed by PEC nanowires. Relying on an eigenwave expansion formalism, it was shown that in the absence of material dispersion the stopping power is an increasing linear function the charged particles velocity. In addition, it was explained how the theoretical framework can be generalized to dispersive media. The results are completely consistent with the numerical analysis reported in [18], and provide further physical insights of the mechanisms that enable the threshold-free Cherenkov radiation by the uniaxial wire medium.

**Acknowledgement:** This work was funded by Fundação para a Ciência e a Tecnologia under project PTDC/EEI-TEL/2764/2012. T. A. Morgado and D. E. Fernandes acknowledge financial support by Fundação para a Ciência e a Tecnologia (FCT/POPH) and the cofinancing of Fundo Social Europeu under the Post-Doctoral fellowship SFRH/BPD/84467/2012 and Doctoral fellowship SFRH/BD/70893/2010, respectively.

## APPENDIX A

In this Appendix, it is shown how the system of equations that define the electrodynamics of the uniaxial wire medium can be written in the compact form as $\hat{L}\mathbf{F} = i\overline{\overline{\mathbf{M}}} \cdot (\partial \mathbf{F}/\partial t)$, where $\mathbf{F} = (\mathbf{E}, \mathbf{H}, \varphi, I)^T$ is the eight-component state vector.

Within the quasi-static homogenization approach introduced in [25], the macroscopic electromagnetic fields in a uniaxial wire medium formed by PEC wires satisfy the following equations:

$$\nabla \times \mathbf{E} = -\mu_0 \frac{\partial \mathbf{H}}{\partial t}, \tag{A1}$$

$$\nabla \times \mathbf{H} = \varepsilon_0 \varepsilon_\mathrm{h} \frac{\partial \mathbf{E}}{\partial t} + \frac{I}{A_\mathrm{c}} \hat{\mathbf{z}}, \tag{A2}$$

$$\frac{\partial I}{\partial z} = -C_\mathrm{w} \frac{\partial \varphi}{\partial t}, \tag{A3}$$

$$\frac{\partial \varphi}{\partial z} = -L_\text{w} \frac{\partial I}{\partial t} + E_z. \tag{A4}$$

After simple manipulations, it is possible to rewrite this system of equations in the form $\hat{L}\mathbf{F} = i\overline{\overline{\mathbf{M}}} \cdot (\partial \mathbf{F}/\partial t)$, where the first-order linear differential operator $\hat{L}$ is

$$\hat{L} = \begin{pmatrix} 0 & i\nabla\times & 0 & \dfrac{-i}{A_c}\hat{\mathbf{z}} \\ -i\nabla\times & 0 & 0 & 0 \\ 0 & 0 & 0 & \dfrac{-i}{A_c}\dfrac{\partial}{\partial z} \\ \dfrac{i}{A_c}\hat{\mathbf{z}} & 0 & \dfrac{-i}{A_c}\dfrac{\partial}{\partial z} & 0 \end{pmatrix}, \tag{A5}$$

and the material matrix $\overline{\overline{\mathbf{M}}}$ that describes the electromagnetic response of the medium is

$$\overline{\overline{\mathbf{M}}} = \begin{pmatrix} \varepsilon_\text{h}\varepsilon_0 & 0 & 0 & 0 \\ 0 & \mu_0 & 0 & 0 \\ 0 & 0 & \dfrac{C_\text{w}}{A_\text{c}} & 0 \\ 0 & 0 & 0 & \dfrac{L_\text{w}}{A_\text{c}} \end{pmatrix} \tag{A6}$$

The matrix $\overline{\overline{\mathbf{M}}}$ is Hermitian and real valued, i.e. $\overline{\overline{\mathbf{M}}} = \overline{\overline{\mathbf{M}}}^\dagger$.